\documentclass[preprint,a4paper,11pt]{elsarticle}
\usepackage{graphicx}
\usepackage{amssymb}
\usepackage{mathrsfs}
\usepackage{bbm}
\usepackage{ngerman}
\usepackage{geometry}
\usepackage[english]{babel}
\journal{Computer Physics Communications}

\begin{document}
\begin{frontmatter}
\title{Examining the Analytic Structure of Green's Functions:\\ 
Massive Parallel Complex Integration using GPUs}
\author[label1]{Andreas Windisch}
\author[label1]{Reinhard Alkofer}
\author[label2]{Gundolf Haase}
\author[label2]{Manfred Liebmann}
\address[label1]{Institut f\"ur Physik, Universit\"at Graz, 
Universit\"atsplatz 5, 8010 Graz, Austria }
\address[label2]{Institut f\"ur Mathematik und wissenschaftliches Rechnen, 
Heinrichstra\ss e 36, 8010 Graz, Austria }

\begin{abstract} Graphics Processing Units (GPUs) are employed for a numerical
determination  of the analytic structure of two-point correlation functions of
Quantum Field Theories. These functions are represented through integrals in
$d$-dimensional Euclidean momentum space. Such integrals can in general not be
solved analytically, and therefore one has to rely on numerical procedures to
extract their analytic structures if needed. After describing the general
outline of the corresponding algorithm we demonstrate the procedure by providing
a completely worked-out example in four dimensions for which an exact solution
exists. We resolve the analytic structure by highly parallel evaluation of the
correlation functions momentum space integral in the complex plane. The
(logarithmically) divergent integral is regularized by applying a BPHZ-like
Taylor subtraction to the integrand. We find perfect agreement with the exact
solution. The fact that each point in the complex plane does not need any
information from other points makes this a perfect candidate for GPU treatment.
A significant gain in speed as compared to sequential execution is obtained.
We also provide typical running times on several GPUs. 
\end{abstract}

\end{frontmatter}

\section{Introduction}
\label{intro}

In quantum field theory (QFT), the analytic structure of a correlation or Green
function  is tied to the physical interpretation of the propagating degree of
freedom.  Hereby, the analytic structure in the complex plane spanned by
analytic  continuation of the square of the external momentum is considered. 
(NB: We work throughout in Euclidean momentum space, {\it i.e.}, we consider all
integrals after a Wick rotation has been performed.) For example,  a two-point
function of a massive scalar particle features a single pole in  the complex
plane. In  momentum space this pole occurs in the timelike  momentum region at
$p^2=-m^2$, with $p$ being the external momentum and $m$  the mass of the
particle. Furthermore, if there are more than just one particle  making up the
correlator, a branch cut appears starting at a threshold $p^2=-4m^2$.  Also the
occurrence of bound states among the particles forming the correlator  is
evident within the correlator's analytic structure, since then additional  poles
exists below the two-particle threshold in this case. 

Also the property of positivity of a Green function is encoded in its analytic
structure. (More precisely, in the Euclidean case, the analytic structure
determines whether the property  of reflection positivity of the correlator's
Schwinger function is fulfilled.)  Positivity violating Green functions do not
possess a K\"all\'en-Lehmann  representation
\cite{Kallen:1952zz,Lehmann:1954xi}. Their negative norm  contributions do not
allow for a probabilistic interpretation as demanded by a  quantum theory.
Positivity is a necessary condition for a certain state to be  part of the
physical spectrum of asymptotic states. Correspondingly, the investigation of
the  analytic structure has been subject of many preceding studies, see {\it
e.g.}, refs.\  
\cite{Zwanziger:1989mf,Maris:1993pm,Maris:1995ns,Alkofer:2003jj,Baulieu:2009ha}
and references therein. 

The main reason for the development of this code is an investigation of exactly 
this kind \cite{aw3}. There, the field-strength tensor of pure Yang-Mills (YM) 
theory is used to construct the correlator
\begin{equation}
\langle F^2(x)F^2(0)\rangle=\int\frac{d^Dp}{(2\pi)^D}\exp\{i p\cdot x\}
\mathscr{G}(p^2),
\label{1}
\end{equation}  
where $\mathscr{G}(p^2)$ is the corresponding (Euclidean) momentum space 
operator whose analytic structure is to be investigated, and $D$ is the number
of space-time dimensions. Here we have already exploited the fact that due to
Poincar\'e invariance of YM theory $\mathscr{G}(p^2)$ depends only on the
length of the four-vector $p$, {\it i.e.}, on $p^2$. 
The YM field-strength tensor is given by 
\begin{equation}
F_{\mu\nu}^a=\partial_\mu A_\nu^a-\partial_\nu A_\mu^a+gf^{abc}A_\mu^bA_\nu^c,
\label{2}
\end{equation}
and the square entering the correlator reads (using Einstein's sum convention)
\begin{equation}
F^2=F_{\mu\nu}^aF_{\mu\nu}^a.
\label{3}
\end{equation}
YM theory accounts for gauge field dynamics only, thus only gluonic degrees of 
freedom are considered. The gauge fields entering the correlator (\ref{1}) 
might form a bound state, since the non-Abelian character of YM-theory induces 
self-interactions among the gluons. Such bound states are called glueballs
(see, {\it e.g.}, ref.\ \cite{Mathieu:2008me}). In a (hypothetical) exact
calculation they will lead to poles in the expression (\ref{1}). 
Since gluon propagators of different kinds are used as an input in ref.\ 
\cite{aw3}, 
we developed this code which performs the investigation of the analytic 
structure numerically.  Note that such integral expressions cannot be solved in
general with  conventional methods.

This article is intended to provide a step--by--step tutorial for performing 
such an investigation numerically, using the power of parallelism provided by 
Graphics Processing Units. The code has been developed following the 
\texttt{FORTRAN90} standard and has been extended to GPUs by using CUDA Fortran, 
provided by the Portland Group \cite{PGI,PGI2}. Since each evaluation point in 
the complex plane can be treated on its own, the problem is perfectly suitable 
for a GPU treatment.

This article is organized as follows: Sect.\ \ref{formulation} provides the 
guideline how to extract the analytic structure of a correlator numerically, 
while Sect.\ \ref{cuda} presents a short introduction to GPU computing using 
CUDA \cite{NVIDIA} to establish the terminology. 
The numerical implementation of the procedure is described in Sect.\ 
\ref{numerics}. A worked example which can be solved analytically serves as a 
test case in Sect.\ \ref{test}, where we follow and detail the procedure given
in Sect.\ \ref{formulation} to reproduce the exact results numerically. 
We conclude in Section \ref{conclusions}.

\section{\label{formulation}{Step--by--Step to the solution}}

Let us assume we have the momentum space operator $\mathscr{G}(p^2)$ 
corresponding to a certain two-point function in four-dimensional Euclidean 
space as a starting point. Note that the procedure can be applied to arbitrary 
dimensions by trivial modifications, but let us restrict to the physical four 
dimensions for simplicity. The generic structure of such an operator expressed 
through an integral over the internal momentum is
\begin{equation}
\mathscr{G}(p^2)=\int_{\mathbbm{R}^4}\frac{d^4k}{(2\pi)^4}
\mathscr{F}(p^2,k^2,p\cdot k),
\label{4}
\end{equation} 
where $\mathscr{F}(p^2,k^2,p\cdot k)$ is the corresponding integrand which is  a
function of the squares of the external and internal momenta ($p^2$ and $k^2$,
respectively), as well as of the scalar product between the external and the 
internal momentum. The aim is to find the analytic structure of 
$\mathscr{G}(p^2)$ in the complex $p^2$-plane numerically.  In the following we
provide the steps that one has to carry out to achieve  this goal. All one has
to provide to follow this procedure is an operator  expressed in the form as
denoted in eq.\ (\ref{4}).  The steps presented below are then used in the
worked example in Sect.\  \ref{test}.  Steps which are labeled with the
attribute (A) are to be performed analytical, while steps carrying the attribute
(N) are numerical steps. One step carries the attribute (A,N) because one can
choose to work it out either analytical or numerically. In the worked example we
show both possibilities.

\begin{itemize}
\item \textbf{STEP 1, (A): Express (\ref{4}) in hyperspherical coordinates}\\
For a subsequent numerical treatment it is convenient to transform the 
integral in eq.\ (\ref{4}) into hyperspherical coordinates: 
\begin{equation}
\label{5}
\int_{\mathbbm{R}^4}d^4k\rightarrow \int_0^{2\pi}d\phi\ \int_0^\infty dk \, 
k^3\  \int_0^\pi d\theta_1\sin^2\theta_1\ \int_0^\pi d\theta_2\sin\theta_2.
\end{equation}
Applying another change of variables by introducing
\begin{eqnarray}
\label{6}
y\equiv k^2, &\rightarrow& dy=2k dk,\\
\label{7}
\theta_1\equiv \arccos z, &\rightarrow& d\theta_1=-\frac{1}{\sqrt{1-z^2}}dz,\\
\label{8}
\theta_2\equiv \arccos w, &\rightarrow& d\theta_2=-\frac{1}{\sqrt{1-w^2}}dz,
\end{eqnarray}
as well as relabeling $p^2\rightarrow x$ after the transformation, 
we arrive at
\begin{equation}
\label{9}
\int_{\mathbbm{R}^4}d^4k\rightarrow \frac{1}{2}\int_0^{2\pi}d\phi\ 
\int_0^\infty dyy\ \int_{-1}^1 dz\sqrt{1-z^2}\int_{-1}^1 dw.
\end{equation}
Rewriting the integrand of (\ref{4}) in the new variables we see that the 
integrand depends on $x,\ y$ and $z$  only,
\begin{equation}
\label{10}
\mathscr{F}(p^2,k^2,|p||k|\cos \theta_1)\rightarrow 
\mathscr{F}(x,y,\sqrt{x}\sqrt{y}\ z).
\end{equation}    
Thus we can integrate $\phi$ and $w$ trivially:
\begin{equation}
\label{11}
\mathscr{G}(x)=\frac{1}{(2\pi)^3}\int_0^\infty dy y\int_{-1}^1 dz\sqrt{1-z^2}
\mathscr{F}(x,y,\sqrt{x}\sqrt{y}\ z).
\end{equation}

\item \textbf{STEP 2, (A): Regularization (if applicable)}\\
In four dimension an integral like expression (\ref{4}) typically diverges
logarithmically. If this is the case one has to regularize the integral.
This step can be skipped if the integral is already finite.
Here we use the BPHZ-procedure
\cite{Bogoliubov:1957gp,Bogolyubov:1980nc,Hepp:1966eg,Zimmermann:1969jj} to
obtain a finite expression. The so-called superficial degree of divergence (SDD)
can be determined by counting the powers of the inner momenta that appear in
$\mathscr{F}(x,y,\sqrt{x}\sqrt{y}\ z)$, as well as the powers of the inner momenta 
present due to the integral measure. Let the SDD $=n$. Then we can render the
integral finite by replacing the integrand $\mathscr{F}(x,y,\sqrt{x}\sqrt{y}\
z)$ with $\mathscr{F}_{sub}(x,y,\sqrt{x}\sqrt{y}\ z)$, which is obtained by
applying a Taylor-subtraction operator $t^n$ up to the order of $n$ to the
initial integrand,
\begin{equation}
\label{12}
\mathscr{F}_{sub}(x,y,\sqrt{x}\sqrt{y}\ z)=(1-t^n)
\mathscr{F}(x,y,\sqrt{x}\sqrt{y}\ z),
\end{equation}  
where $t^n$ is given by
\begin{equation}
\label{13}
t^n=\sum_{i=0}^n\frac{x^i}{i!}
\left[\frac{\partial^i}{\partial x^i}\right]_{x=0}.
\end{equation}
Performing this step, we arrive at
\begin{equation}
\label{14}
\mathscr{G}_{Sub}(x)=\frac{1}{(2\pi)^3}\int_0^{\Lambda^2} dy y
\int_{-1}^1 dz\sqrt{1-z^2}\mathscr{F}_{Sub}(x,y,\sqrt{x}\sqrt{y}\ z),
\end{equation} 
where upper integration limit $\Lambda^2$, introduced purely for numerical
reasons, assumes some large finite value. Note that the BPHZ procedure 
ensures cut-off independence, {\it i.e.,} 
the limit $\lim_{\Lambda^2}\to\infty$ does not alter the value of the integral.

\item \textbf{STEP 3, (A,N): Analytic continuation}\\
As it has been pointed out in ref.\ \cite{Alkofer:2003jj}, the $z$-integral 
induces a branch cut in the complex $y$-plane. This happens when one  integrates
over the (integrable) singularities of the $z$-integrand.  For certain values of
$x$, the branch cut in the $y$-plane might obstruct  the contour of the
$y$-integration along the positive real axis, as it is used in equation
(\ref{14}). If the integrand of the angular ($z$) integral is a well behaved
function, and if there are no poles of the remaining integrand for the
$y$-integral on the positive real $y$-axis, then there is nothing to do and this
step can be skipped. Usually, the region in the complex $x$-plane for which the
induced branch cut does not interfere with the $y$-contour along the positive
real axis is rather small. To avoid troubles, one has to perform the angular
integration (either analytical or numerically) and look at the results in the
complex $y$-plane. This can also be achieved by finding the complex $y$-values
for which the $z$-integrand for a given complex $x$ becomes singular. In the
usual case of the integrand being a fraction, this simply reduces to putting the
denominator polynomial to zero and solve this equation with respect to (complex)
$y$ for a certain (complex) value of $x$.\par As soon as this troublesome
regions in the complex $x$-plane have been identified, one has to deform the
$y$-contour in the complex $y$-plane accordingly, avoiding the branch-cut
present due to the angular integration, as well as poles which might be present
due to the $y$-integrand itself. The most accurate form of the integral
corresponding to $\mathscr{G}_{Sub}(x)$ is   
\begin{equation}
\label{15}
\mathscr{G}_{Sub}(x)=\frac{1}{(2\pi)^3}\int_\mathscr{C} dy y\int_{-1}^1 dz
\sqrt{1-z^2}\mathscr{F}_{Sub}(x,y,\sqrt{x}\sqrt{y}\ z),
\end{equation} 
with $\mathscr{C}$ being the deformed contour connecting $0$ with 
$\Lambda^2$, while avoiding the cut and the poles.  

\item \textbf{STEP 4, (N): Preparation}\\
Now we have done all the analytical work involved in this investigation. The
first (numerical) step is to determine a $(M\times N)$ matrix $X$ which holds
the $x$ values at which we want to evaluate $\mathscr{G}_{Sub}(x)$ as given by
equation (\ref{15}), with the contour adjusted as necessary (see \textbf{STEP
3}). $M$ denotes the number of points along the real axis of $x$, and $N$ is the
number of points along the imaginary axis of $x$. The limits for the complex
$x$-values have to be chosen as desired. The region of interest in the complex
$x$-plane is then discretized and mapped to a matrix $X$. This step is already
performed on the GPU (see Section \ref{numerics} for the implementation of the
numerics).    

\item \textbf{STEP 5, (N): Evaluation of the integrals}\\
This is the final step in this procedure. It is also performed on the GPU, using
its parallelism capabilities. For each entry of the matrix $X$ we have to
evaluate equation (\ref{15}), which we do by using non-adaptive quadrature rules
for approximating the values of the integrals. On the GPU we perform these
integrations in parallel for a $(m\times n)$ sub-matrix of $X$, where the
maximal size of the sub-matrix depends on the architecture of the GPU. The
result is stored to a file, the subsequent graphics processing of the data is
then performed by using \texttt{Mathematica} \cite{Mathematica8}.    

\end{itemize}

This concludes our generic discussion of the procedure. The numerical steps
(\textbf{STEP 3, STEP 4} and \textbf{STEP 5}) are discussed in detail in Sect.\
\ref{numerics}. The whole procedure (\textbf{STEP 1} - \textbf{STEP 5}) is
demonstrated in great detail by the worked example in Sect.\ \ref{test}.

\section{\label{cuda}{Introduction to CUDA}}

In this section we briefly review some features of CUDA in order
to make this article self-contained. We only describe such features we will
need in the following section, for a  more detailed description the reader is
referred to ref.\ \cite{NVIDIA}.

Compute Unified Device Architecture (CUDA) is a general purpose
parallel computing architecture provided by NVIDIA \cite{NVIDIA}. CUDA-enabled
devices can be addressed using C (with some extensions) as a programming
language, but there are also other high-level languages available.
This study has been performed by implementing the code using CUDA Fortran,
provided by PGI \cite{PGI,PGI2}. CUDA Fortran is a small set of extensions to
Fortran that allows one to use the computing resources of CUDA-enabled devices.
A brief review of the realm of CUDA is in order before we discuss the actual
implementation of the steps described in the previous section.\par In contrast
to CPUs, GPUs are devices specialized in heavy-duty parallel calculations, as
they are demanded by their primary purpose, i.e., 3D rendering. A GPU features
several SIMD multiprocessors. The CUDA programming model is
organized as follows. The code controlling the calculation runs as a sequential
code on a CPU called the \textit{host}. The task of the host part is to control
the flow of data to and from the GPU, which is called the \textit{device}. Each
CUDA program consist thus of a host part and a device part. A function that is
executed on the GPU is called a \textit{kernel}. 
CUDA possesses a 3-level thread hierarchy. The lowest
level of the model is a \textit{thread}. The threads are executed on a scalable
array of multithreaded \textit{Streaming Processors} (SMs). A multiprocessor 
uses the \textit{Single-Instruction Multiple-Thread} (SIMT) architecture to perform
the concurrent execution of the threads. The multiprocessor partitions the threads
in groups of 32 threads, called \textit{warps}, which are then scheduled and executed. 
One common instruction is executed within a warp at a given time.The next element in 
the hierarchy is called a \textit{block}. Each block contains a certain number of
threads, which should be a multiple of the warp-size for an efficient usage of
the hardware. The maximum number of threads per block is limited and determined
by the hardware in use. The highest level in the hierarchy is called the
\textit{grid}. The grid consists of blocks and is assigned to a device. On the
device, the blocks within the grid are distributed among the SMs, where subsets
of the blocks, the warps, are scheduled. The threads of the warp are then
executed on the cores of the SM. Blocks and grids can be either 1, 2 or 3-dimensional,
which also depends on the hardware. Each thread within the hierarchy
possesses a unique identification number, the thread-ID, which can be calculated
from the position of the block in the grid and the position of the thread in the
block. Besides the thread hierarchy, there exists a memory hierarchy on the
device. The \textit{global device memory} can be accessed by each thread, where
each thread can read and write on this memory. However, it is not a small
latency memory. In the so-called Fermi architecture, each SM possesses a on-chip 
memory of 64 KB. One can either configure the on-chip memory to use 48 KB for shared
memory with 16 KB L1 cache or to use 16 KB for shared memory with 48 KB L1 cache. 
Fermi furthermore features a 768 KB L2 cache which connects all SMs.

A CUDA program is usually of the following form. Parts running on the host are
labeled with the attribute (H), while device parts are labeled with (D).

\begin{itemize}
\item \textbf{Initialization (H)}
The program starts with a host part, where data is allocated in the host memory,
as well as in the device memory. The grid- and blocksize and dimensions are
determined, variables are initialized and prepared for the parallel execution on
the device. Data is transferred from the host to the device.

\item \textbf{Kernel execution (D)}
One or more kernels are executed on the device by a kernel call of the host.
Each kernel has to be called together with its grid- and blocksize, which also
includes the information about their dimensionality. The blocks of the grid of
the kernel are executed on the SMs of the device.

\item \textbf{Finalization (H)}
Once the device executions have finished, the result, which still resides on the
device, has to be transferred back to the host where it can be stored to file.
Finalizing steps such as memory deallocation are performed and the program
closes.

\end{itemize} 

This rather brief discussion covers only the very basics of the programing 
model, but should provide enough understanding to follow the discussion in the 
next section. 

\section{\label{numerics}{Numerical Implementation}}

On the basis of the short introduction to CUDA outlined in the previous section,
we can now introduce the actual implementation of the numerical steps of Section
\ref{formulation}. The centerpiece of course is the integration, which we
perform by numerical quadrature. In particular we use Gauss-Legendre quadrature
\cite{Abramowitz:1964as} for the radial ($y$) integral, and Tanh-Sinh quadrature
(also known as double exponential quadrature) \cite{Takahasi:1974tm} for the
angular ($z$) integral. We also implemented Gauss-Chebyshev quadratures
\cite{Abramowitz:1964as} (by using both, polynomials of the first and second
kind, which we use as convenient depending on the dimension the problem is
located in) to have a check on the angular integration results. The code is
organized as follows, where again we make use of the labels (H) for code running
on the host and (D) for device code respectively.

\begin{itemize}
\item \textbf{Initialization (H)}\\
Several preparation steps are necessary.
\begin{itemize}
\item \textbf{Allocation}\\
Allocation on both, the host and the device is performed for the weights and
nodes needed for the quadratures. Since it is necessary to integrate along
contours which are running close to singular structures in the complex plane, we
have to use a huge number of nodes to get a smooth picture in the end. Another
option would have been to use adaptive strategies, but they tend to get stuck
near singularities and are more complicated to implement. We achieved good
results with this procedure.

\item \textbf{Preparation}\\
The calculation of the weights and nodes that are to be stored in the arrays of
size $ngl$ for Gauss-Legendre and $ndbl$ for Tanh-Sinh quadrature respectively,
are calculated on the host. In well-behaved areas it is by far sufficient to use
$ngl=64$ and $ndbl=49$. However, in regions where the contour comes close to the
singular structures living in the complex $y$-plane, one only obtains good
results with much more nodes. Note furthermore, that while we only need one
array for the angular nodes, we need $n$ arrays for the radial nodes, where $n$
is the number of the sub-contours needed to form the contour $\mathscr{C}$ that
connects $0$ with $\Lambda^2$ in the complex $y$-plane. Depending on the
parametrization, we have $n=1$ up to $n=5$ in our worked example. Besides the
arrays for the weights and nodes, we need two $M\times N$ matrices of the
data-type complex. The matrix $X$ will contain the complex $x$ values at which
we intend to perform the calculation, the matrix $M$ is used to store the result
in the end. Furthermore, the grid and block dimension is set.   

\item \textbf{Transfer}\\
The data that has been calculated on the host is transferred to the device
global memory. 

\end{itemize}

\item \textbf{Kernel 1 (D): Discretize the region of evaluation}\\
This kernel corresponds to \textbf{STEP 4} of Section \ref{formulation}. This
kernel is used to fill the matrix $X$ with the complex values of the region in
the complex $x$-plane where $\mathscr{G}_{Sub}(x)$ is to be evaluated. We used
square matrices with $128^2$ points. Usually, $128^2$ points are absolutely
sufficient to produce a reasonable picture of the area, for some purposes we
also used $256^2$ and $512^2$ points. We usually restrict the real and imaginary
part of $x$ to range between $\pm 5$. The kernel is called with blocks of
$16\times 16$ in size, the grid in case of the $128^2$ matrix is then formed by
$8\times 8$ blocks. Each thread, which is described by a unique tuple $(i,j)$
writes the complex number of $x$ to the matrix element $X(i,j)$, where it
determines the complex number by calculating a homogeneous distribution of the
$M\times N$ points within the given range. The result is a matrix which holds a
discretized version of the region of interest in the complex $x$-plane. 

\item \textbf{Kernel 2 (D): Evaluation}\\
This kernel corresponds to \textbf{STEP 5} of Section \ref{formulation}. It is
the center part of the calculation. The block and grid sizes are the same as for
Kernel~1. Each thread within a block is assigned to exactly one value of the
$X$-matrix and calculates the double integral of equation (\ref{15}), where it
has to choose the contour that has been assigned to the area the point of
evaluation $x_{ij}\in X$ belongs to. In our worked example we had to distinguish 4
areas in the complex $x$-plane, so we had to use 4 different contours. The
decision which contour is the right one can be made by a simple IF-THEN-ELSE
statement. The result produced by each thread corresponding to a certain value
of $x$ is then stored in the result matrix $M$. Once this kernel has finished
its execution, the host takes over again.   

\item \textbf{Finalization}\\
The host transfers the result stored in the matrix $M$ from the device back to
the host and produces a stream that stores the result, together with the complex
$x$ value in a file. The file-name is generated dynamically and consists of
several parameters to identify the run.  

\end{itemize}

The whole procedure will be detailed in the worked example in the 
following section.

\section{\label{test}{A Worked Example}}

The procedure as proposed in the step--by--step recipe of Sect.\
\ref{formulation} is here detailed using as  example the correlator given in
eq.\ (32) of ref.\ \cite{Baulieu:2009ha}. There exists an exact solution for
this correlator which makes this example a perfect test-case for the numerics.
The correlator is given by
\begin{equation}
\label{16}
\mathscr{G}(p^2)=\int\frac{d^Dk}{(2\pi)^D}\frac{1}{(p-k)^2-i\sqrt{2}\theta^2}
\frac{1}{k^2+i\sqrt{2}\theta^2}.
\end{equation}
In ref.\ \cite{Baulieu:2009ha} it has been shown that in four dimensions 
for $2\sqrt{2}\theta^2=1$ the integral in (\ref{16}) can be done analytically
for the  regularized expression yielding
\begin{equation}
\label{17}
\mathscr{G}_{sub}(x)=\frac{1}{16\pi^2}\left(1-\frac{\pi}{2x}+
\frac{\sqrt{1-x^2}}{x}\arccos(x)\right).
\end{equation} 
Our numerical result will be compared to it, however, it is useful to rescale 
(\ref{17}) to get rid of the prefactor:
\begin{equation}
\label{18}
\mathscr{G}_{sub,rescaled}(x)=\left(1-\frac{\pi}{2x}+\frac{\sqrt{1-x^2}}{x}
\arccos(x)\right).
\end{equation} 
The numerical task starts with eq.\ (\ref{16}) and continues with the steps 
outlined in Sect.\ \ref{formulation}. 

\subsection{\label{step1}{Step 1: Transform the integral in hyperspherical 
coordinates}}

We investigate $(16\pi^2)\times \bigl[$ eq.\ (\ref{16}) $\bigr]$ in four
dimensions with $2\sqrt{2}\theta^2\equiv 1$. Step 1 demands to switch to
hyperspherical coordinates, which gives according to eq.\ (\ref{11})
\begin{equation}
\label{19}
\mathscr{G}_{rescaled}(x)=\frac{16\pi^2}{(2\pi)^3}
\int_0^\infty dy y\int_{-1}^1 dz\sqrt{1-z^2}
\frac{1}{(x+y-2\sqrt{x}\sqrt{y}z-\frac{i}{2})}\frac{1}{(y+\frac{i}{2})}.
\end{equation}

\subsection{\label{step2}{Step 2: Regularization}}

We determine the superficial degree of divergence of the integral by
investigating eq.\ (\ref{16}). In four dimensions we have four powers of the
inner momentum $k$ in the numerator due to the integral measure. The denominator
also produces a highest power of 4 in the momentum $k$. Thus the superficial
degree of divergence is 0, which means that the integral diverges
logarithmically. Following the procedure, we use eq.\ (\ref{12}) with the
definition (\ref{13}) to get the regularized integrand, which is in our case
\begin{eqnarray}
\label{20}
\mathscr{F}_{sub}(x,y,\sqrt{x}\sqrt{y}z)&&=(1-t^0)
\mathscr{F}(x,y,\sqrt{x}\sqrt{y}z)\\
&&=\frac{1}{(x+y-2\sqrt{x}\sqrt{y}z-\frac{i}{2})}
\frac{1}{(y+\frac{i}{2})}-\frac{1}{(y-\frac{i}{2})}\frac{1}{(y+\frac{i}{2})}
\nonumber\\
&&=\frac{-x+2\sqrt{x}\sqrt{y}z}
{(x+y-2\sqrt{x}\sqrt{y}z-\frac{i}{2})(y^2+\frac{1}{4})}.\nonumber
\end{eqnarray} 
Performing the cancellations in the prefactor of (\ref{19}), as well as 
plugging eq.\ (\ref{20}) into  (\ref{14}), the following equation remains,
\begin{equation}
\label{21}
\mathscr{G}_{sub,rescaled}(x)=
\frac{2}{\pi}\int_0^\infty dy y\int_{-1}^1 dz\sqrt{1-z^2}
\frac{-x+2\sqrt{x}\sqrt{y}z}
{(x+y-2\sqrt{x}\sqrt{y}z-\frac{i}{2})(y^2+\frac{1}{4})}.
\end{equation}

\subsection{\label{step3}{Step 3: Analytic continuation}}
\begin{figure}
\centering
\includegraphics[width=13cm]{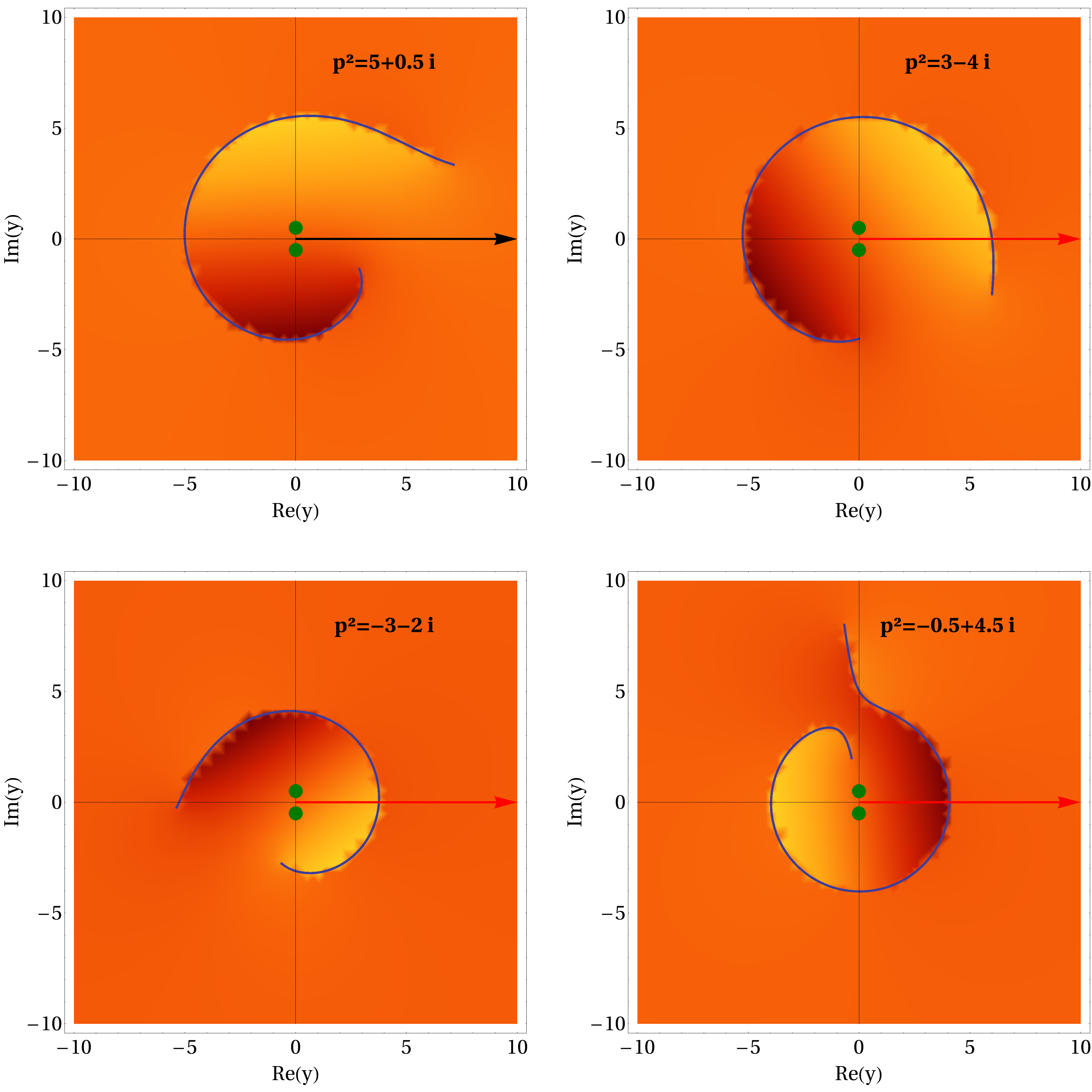}
\caption{Snapshot of the complex $y$-plane after the angular integration for 
four different values of $x=p^2$. The blue curves given by eq.\ (\ref{25}) 
coincide with the rough numerical estimate for the angular integration which 
is expressed by the density-plot in the background. At each point in complex 
$y$-space the branch cut looks different. The original integration contour 
is shown by the arrow. Only in the case shown in the upper left corner we can 
keep the original contour. In all other cases we are blocked by the branch cut.
 The green dots represent the poles present due to term $A$ in eq.\ (\ref{22}).}
\label{fig1}
\end{figure}

\begin{figure}
\centering
\includegraphics[width=10cm]{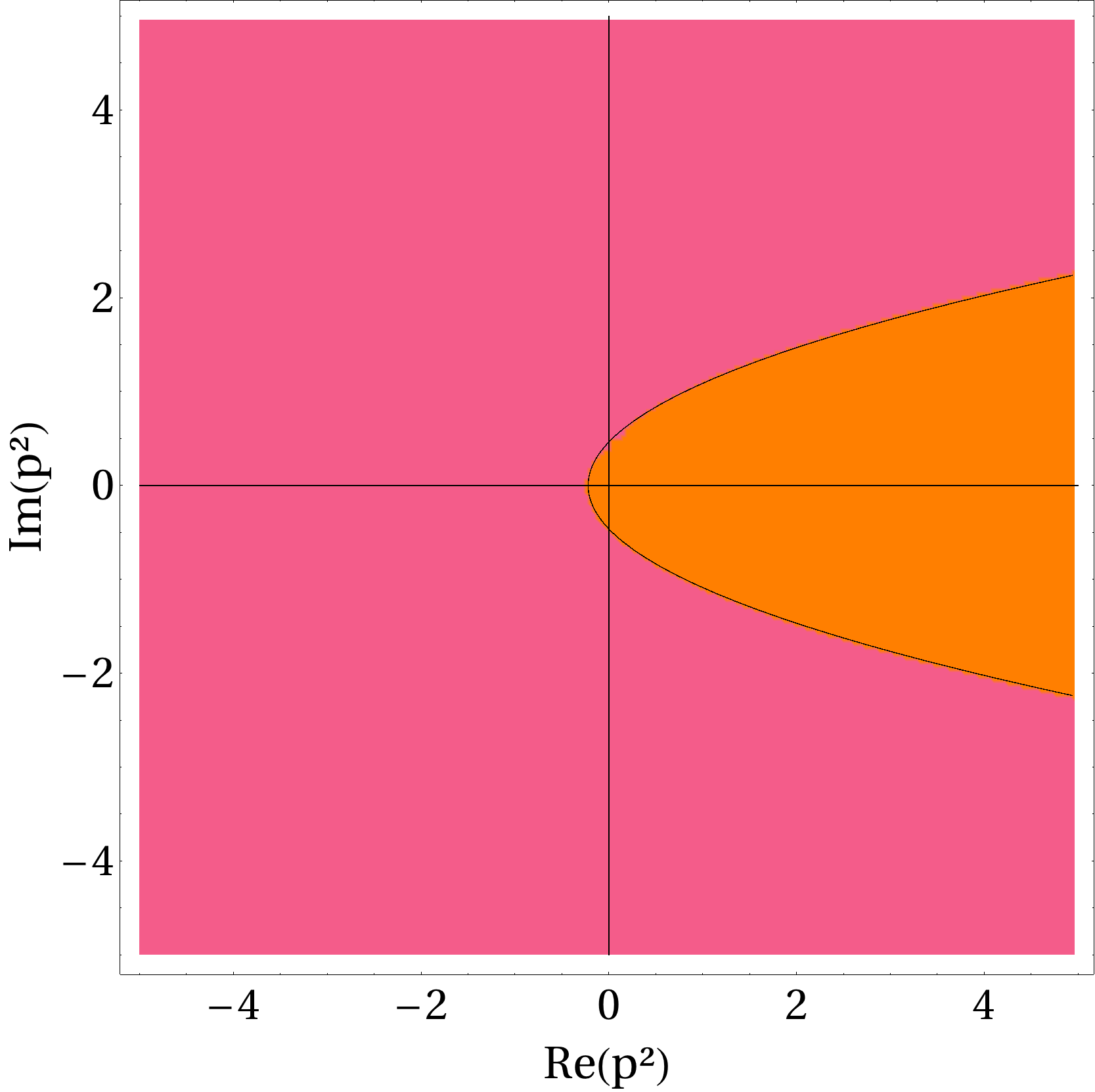}
\caption{The parabola-shaped region on the right is the set of all complex 
$x$-points within the region of evaluation where the branch cut does not 
obstruct the integration contour along the positive real $y$-axis. 
In all other regions we have to deform the contour in an adequate way.}
\label{fig2}
\end{figure}

\begin{figure}
\centering
\includegraphics[width=10cm]{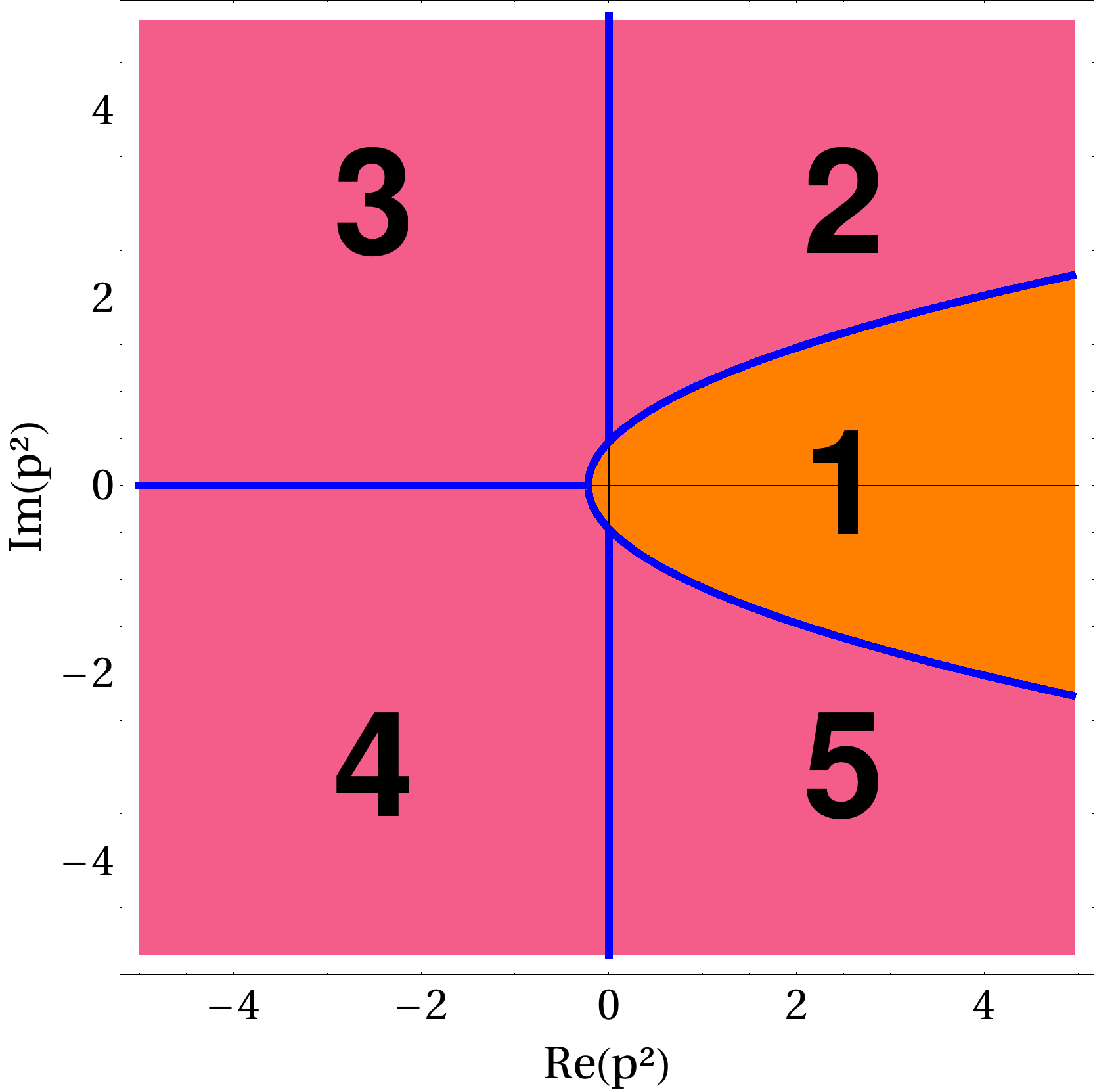}
\caption{We split the complex $x$-plane into 5 regions where we will apply 
different deformations of the integration contour. In region 1 we can keep the 
original contour along the positive real $y$-axis. Furthermore, in the 
regions 2 and 5 we can apply the same contour, such that we are left with 
3 different parametrizations covering the regions 2, 3, 4 and 5, and the 
original contour which we can keep in region 1.}
\label{fig3}
\end{figure}

\begin{figure}
\centering
\includegraphics[width=12cm]{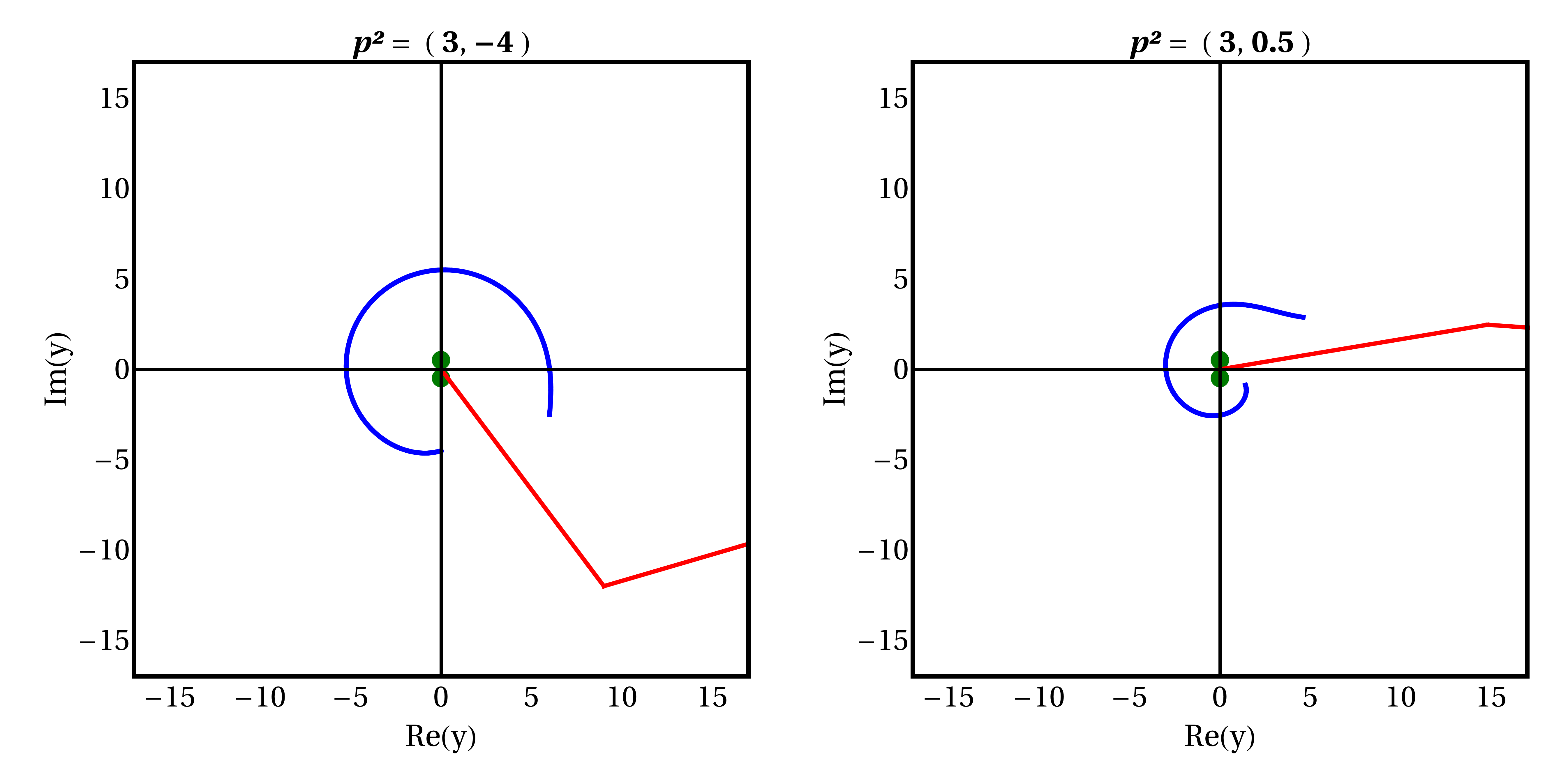}
\caption{The left picture shows the contour as chosen for the regions 2 and 5. 
In the right picture there is no obstruction by the cut and one could in 
principle also use the original contour.}
\label{fig4}
\end{figure}

\begin{figure}
\centering
\includegraphics[width=12cm]{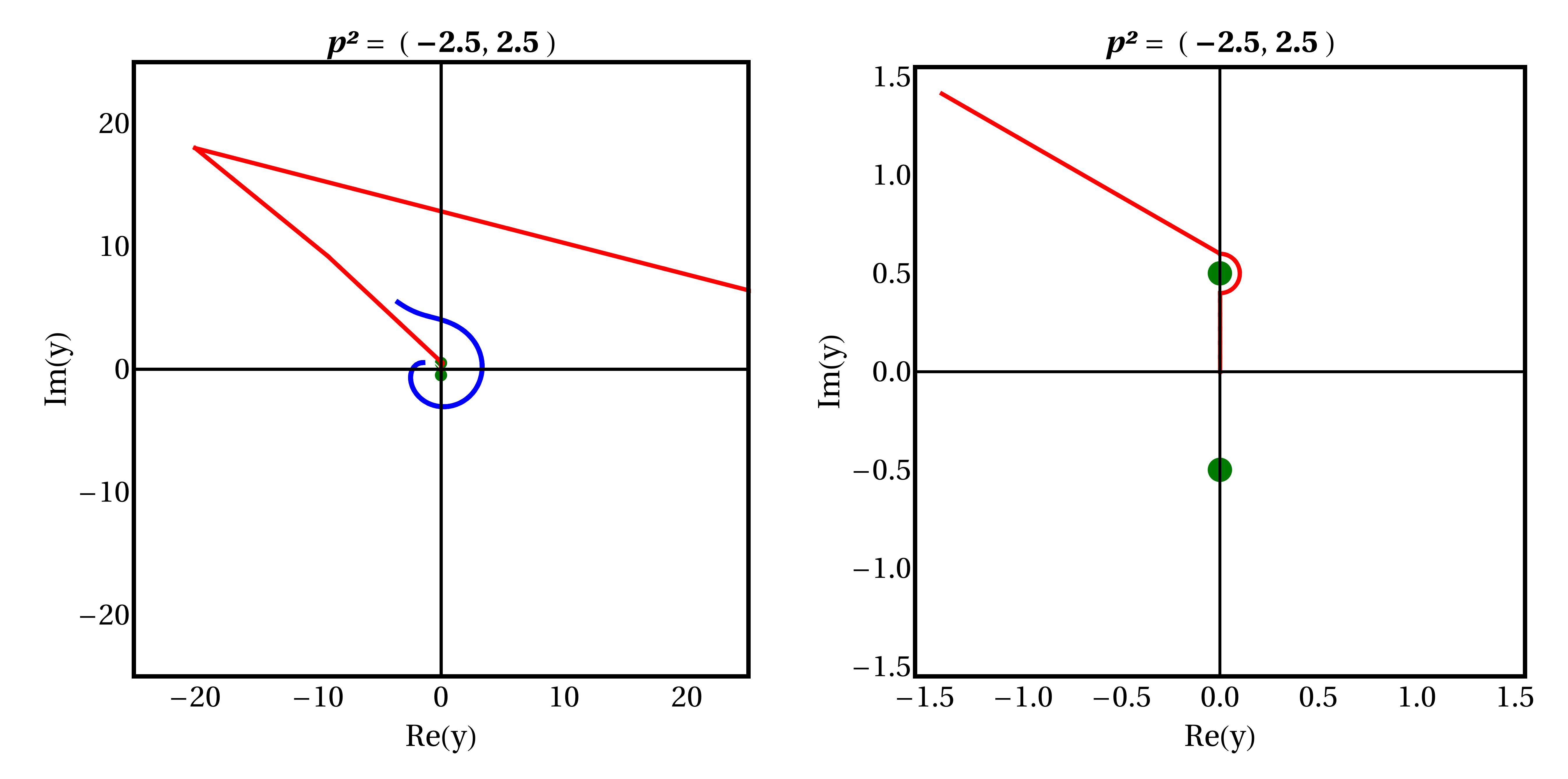}
\caption{The left picture shows the contour as chosen for region 3, the right 
picture shows a close-up. One has to avoid the pole when deforming the contour. 
The contour is closed via the upper half-plane.}
\label{fig5}
\end{figure}

\begin{figure}
\centering
\includegraphics[width=12cm]{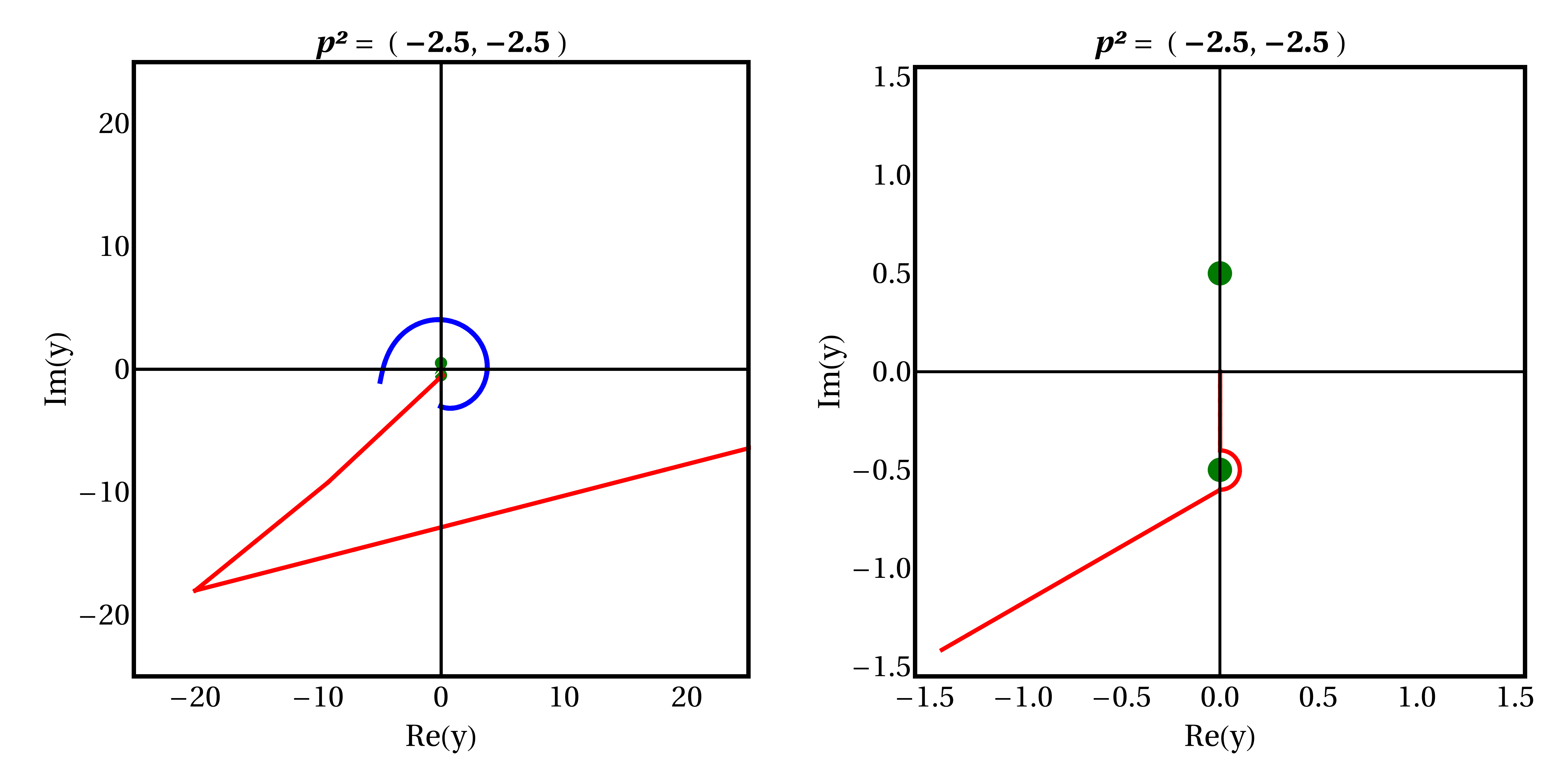}
\caption{The left picture shows the contour as chosen for region 4, 
the right picture shows a close-up. One has to avoid the pole when 
deforming the contour. The contour is closed via the lower half-plane.}
\label{fig6}
\end{figure}

In this step we have to investigate the analytic structure arising in the
complex $y$-plane due to the analytic continuation of $x=p^2$ appearing in the
integrand of the angular integral. The integral is
\begin{equation}
\label{22}
\mathscr{G}_{sub,rescaled}(x)=
\frac{2}{\pi}\int_0^\infty dy 
\underbrace{\frac{y}{y^2+\frac{1}{4}}}_{\equiv A}
\underbrace{\int_{-1}^1 dz\sqrt{1-z^2}\frac{-x+2\sqrt{x}\sqrt{y}z}
{(x+y-2\sqrt{x}\sqrt{y}z-\frac{i}{2})}}_{\equiv B}.
\end{equation}
Clearly, term $A$ induces two poles in the complex $y$-plane, appearing at
$y=\pm\frac{i}{2}$. The angular integral $B$ produces a branch cut in the
complex $y$ plane once the outer momentum $x=p^2$ has been continued to a
complex value. The branch cut at a given value of $x$ in the complex $y$-plane
can be found analytically by finding the poles of the integrand of $B$. In order
to obtain a parametrization, we take the denominator polynomial prior to the
change of variables, {\it i.e.,} we solve
\begin{equation}
\label{23}
p^2+k^2-2pk\cos\theta_1-\frac{i}{2}=0
\end{equation}
with respect to k. The two (redundant) solutions can be written as
\begin{equation}
\label{24}
\tilde\xi(p^2,\theta_1)=p\cos\theta_1\pm\sqrt{-p^2\sin\theta_1+\frac{i}{2}},
\quad 0\leq\theta_1\leq\pi,
\end{equation}
where we named the complex function $\tilde\xi$ instead of $k$ to stress 
that we use it as a parametrization for the branch cut. 
Since we solved eq.\ (\ref{23}) with respect to $k$ for convenience, 
the branch cut in the complex $y=k^2$ plane is then given by 
\begin{equation}
\label{25}
\xi(p^2,\theta_1)=
\left(\sqrt{p^2}\cos\theta_1\pm\sqrt{-p^2\sin\theta_1+\frac{i}{2}}\right)^2,
\quad 0\leq\theta_1\leq\pi.
\end{equation}
With the help of eq.\ (\ref{25}), we can plot the branch cut in the complex
$y$-plane for a given value of $p^2=x$. We also performed this investigation
numerically by simply solving integral $B$ for a given value of $p^2=x$ and for
complex $y$ with \texttt{Mathematica} \cite{Mathematica8}, as well as by using
CUDA-Fortran and a separate kernel to obtain and verify the information. Fig.\
\ref{fig1} shows the complex $y$-plane for some values of $p^2=x$.
In fact, there are not many points in the complex $x$-plane where the branch cut
does not interfere with the original integration contour along the positive real
$y$-axis. We calculated the region where no obstruction occurs numerically by
using a kernel designed for this purpose. Fig.\ \ref{fig2} shows the regions in
the complex $x$-plane where the original contour remains unharmed.
Since the branch cut changes its orientation, size and shape we have to divide
the complex $x$-plane into regions within we can apply the same contour. Fig.\
\ref{fig3} shows the regions we have chosen.
There is nothing to do in region 1, thus we proceed by investigating the other 
regions. In the following we show the parametrizations of the deformed 
contours for the regions 2, 3, 4 and 5. The Figs.\ \ref{fig4}, \ref{fig5} and 
\ref{fig6} show the contours obtained by these parametrizations.
\begin{itemize}
\item \textbf{Region 2 and 5}
\begin{itemize}
\item $\mathscr{C}_{(2,5),1}:\ 0\leq t\leq 1$\\
$\mathscr{C}_{(2,5),1}(t) = 15 t \exp\{i\arg(x)\}$
\item $\mathscr{C}_{(2,5),2}:\ 1\leq t\leq 2$\\
$\mathscr{C}_{(2,5),2}(t) = 15 (2-t)\exp\{i\arg(x)\}-(1-t)\Lambda^2$
\end{itemize}
\item \textbf{Region 3}
\begin{itemize}
\item $\mathscr{C}_{3,1}:\ 0\leq t\leq 1$\\
$\mathscr{C}_{3,1}(t) =t 0.4 i$
\item $\mathscr{C}_{3,2}:\ 1\leq t\leq 2$\\
$\mathscr{C}_{3,2}(t) = 0.1(\sin((2-t)\pi)+i(\cos((2-t)\pi)+5))$
\item $\mathscr{C}_{3,3}:\ 2\leq t\leq 3$\\
$\mathscr{C}_{3,3}(t) =(3-t)(0.6 i)-(2-t)13\exp\{i\arg(x)\}$
\item $\mathscr{C}_{3,4}:\ 3\leq t\leq 4$\\
$\mathscr{C}_{3,4}(t)=(4-t)13\exp\{i\arg(x)\}-(3-t)(-20+18i)$
\item $\mathscr{C}_{3,5}:\ 4\leq t\leq 5$\\
$\mathscr{C}_{3,5}(t)=(4-t)(-20+18 i)-(3-t)\Lambda^2$
\end{itemize}
\item \textbf{Region 4}
\begin{itemize}
\item $\mathscr{C}_{4,1}:\ 0\leq t\leq 1$\\
$\mathscr{C}_{4,1}(t) =-t 0.4 i$
\item $\mathscr{C}_{4,2}:\ 1\leq t\leq 2$\\
$\mathscr{C}_{4,2}(t) = 0.1(\sin((t-2)\pi-\pi)+i(\cos((t-2)\pi-\pi)-5))$
\item $\mathscr{C}_{4,3}:\ 2\leq t\leq 3$\\
$\mathscr{C}_{4,3}(t) =(3-t)(-0.6 i)-(2-t)13\exp\{i\arg(x)\}$
\item $\mathscr{C}_{4,4}:\ 3\leq t\leq 4$\\
$\mathscr{C}_{4,4}(t)=(4-t)13\exp\{i\arg(x)\}-(3-t)(-20-18i)$
\item $\mathscr{C}_{4,5}:\ 4\leq t\leq 5$\\
$\mathscr{C}_{4,5}(t)=(4-t)(-20-18 i)-(3-t)\Lambda^2$
\end{itemize}
\end{itemize}

\subsection{\label{step4}{Step 4: Preparation}}
In this step we have to choose the size of the matrix to which the complex
$x$-plane is mapped to. In this example we used $M\times N=128\times 128$ for
the size of the matrix $X$, and we are interested in the area
$-5\leq\mathfrak{Re}\ x\leq 5$, $-5\leq\mathfrak{Im}\ x\leq 5$. The matrix is
filled with $128^2$ points which are homogeneously distributed over the region
we restricted this consideration to. The CUDA kernel achieving this is called
with a block size of $16\times 16$ threads and with $8\times 8$ blocks forming
the grid. Each thread in the grid, which can be uniquely addressed by a tuple of
numbers $(i,j)$, operates only on the matrix element $X(i,j)$. The indices $i,\
j$ are running from 1 to 128. Thus, for example the thread identified by the
tuple $(34,117)$ operates only on the matrix entry $X(34,117)$. 

\subsection{\label{step5}{Step 5: Evaluation of the integrals}}
This is the last step of our program. The kernel corresponding to this step is
launched with the same parameters as the one in the step before, that is we
still use $16\times 16$ blocks organized in a $8\times 8$ grid. Each thread
operates only on the matrix entry it is assigned to and performs the
integrations according to its position in complex $x$-space. The decision which
contour a certain thread has to choose is made by a simple IF-THEN-ELSE
construction.

\subsection{\label{result}{Results}}
\begin{figure}
\centering
\includegraphics[width=12cm]{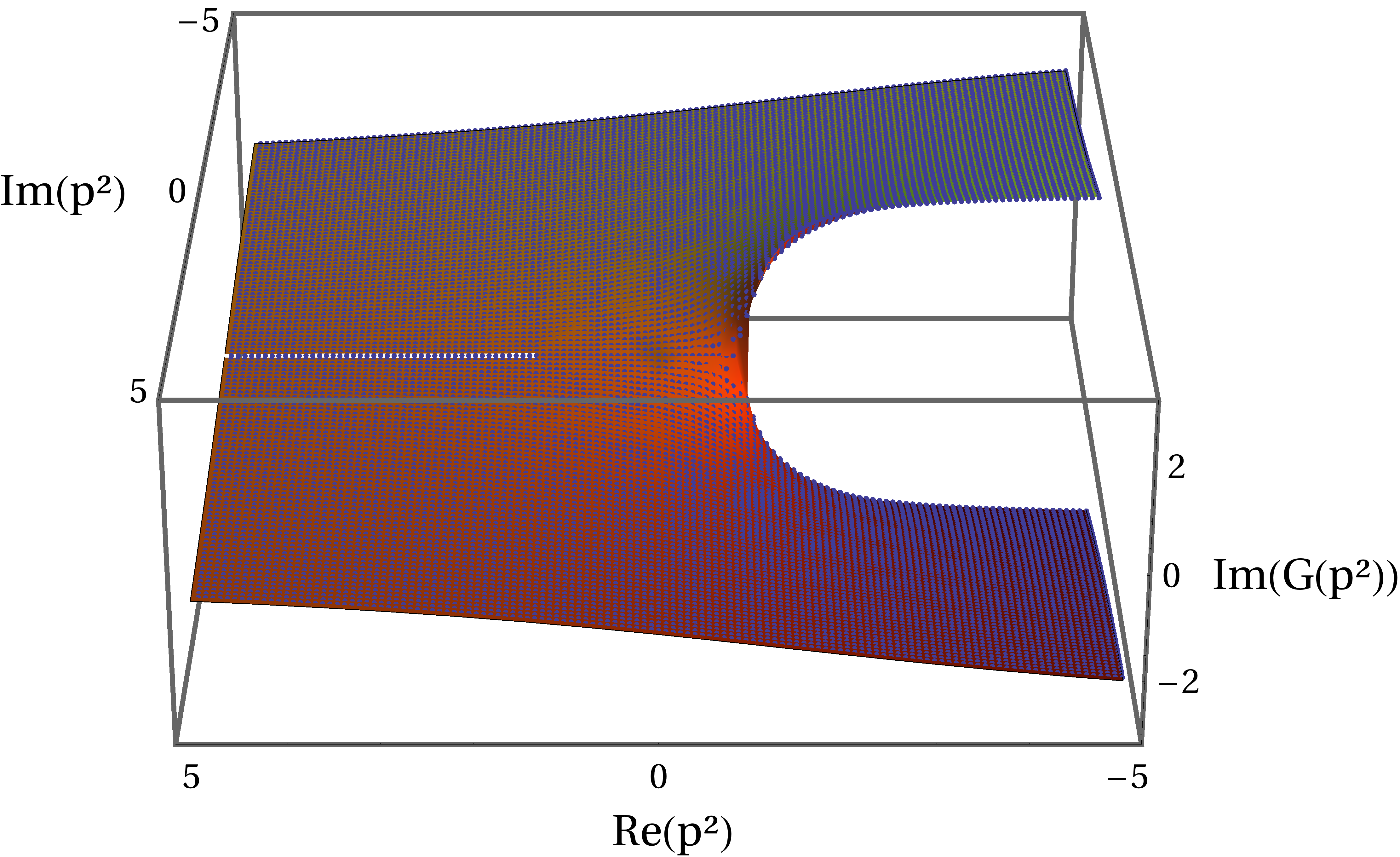}
\caption{The imaginary part of the solution of eq.\ (\ref{16}) after 
regularization and rescaling. The numerical data (blue dots) perfectly 
agrees with the exact solution provided by the surface-plot.}
\label{fig7}
\end{figure}

\begin{figure}
\centering
\includegraphics[width=12cm]{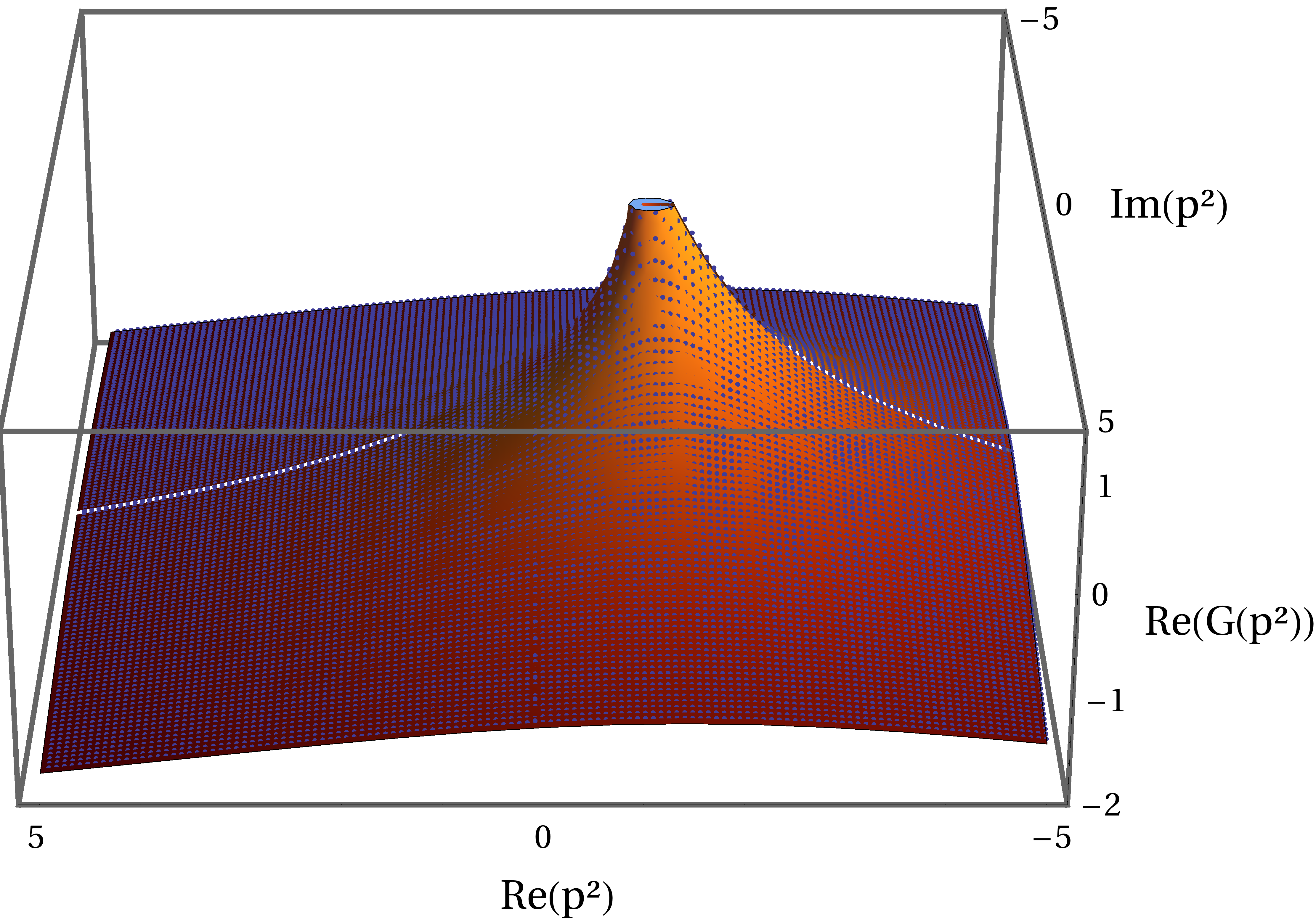}
\caption{The real part of the solution of eq.\ (\ref{16}) after 
regularization and rescaling. The numerical data (blue dots) 
perfectly agrees with the exact solution provided by the surface-plot.}
\label{fig8}
\end{figure}

Finally we can compare the results obtained by numerical integration compared to
the exact solution given by eq.\ (\ref{18}). The plots are shown in the
Figs.\ \ref{fig7} and \ref{fig8}. This concludes our worked example.


\subsection{\label{exec}{Execution time}}
We compared the total execution time of a cheap consumer graphics card and two
high performance graphics cards to the time needed by the code to run on one
core of a modern CPU. The speedup results are presented in Table \ref{tab1}.
\begin{table}
\centering
\begin{tabular}{||c|c|c|c|c||}
\hline
&\textbf{Intel Xeon CPU} &\multicolumn{3}{c||}{\textbf{NVIDIA GeForce GPU}}\\
\hline
Device&X5650 (1 core) & GTX 550 Ti & GTX 480 & Tesla C2070\\
\hline
Runtime& 252m 2s&6m21s&4m21s&2m38s\\
\hline
Speedup&1&$\approx 39.6$&$\approx 57.9$ & $\approx 95.5$\\
\hline
\end{tabular}
\caption{Comparison of running time and speed for a certain set of parameters. 
We used a modern CPU and compared against three different GPUs.}
\label{tab1} 
\end{table}

\section{\label{conclusions}{Conclusions}}

To summarize, we presented the numerical determination of the analytic structure
of correlation functions given by momentum space integrals exploiting the
parallel computing capability of a GPU. This rather complicated numerical
analysis can then be performed within minutes.  We provided a step--by--step
description for the procedure required to obtain a numerical solution, and we
presented a worked example and compared to its exact solution. 

The independence of the points in the complex plane makes such an investigation
a perfect candidate for GPU treatment, which pays off even more with increasing
matrix-size and/or increasing number of nodes. Note that there may be still
potential in speeding up both, the procedure and the code, which we have not
done so far.

The development of this procedure has been the first step in an ongoing
investigation \cite{aw3} of the analytic structure of correlators with 
different expressions as input. Hereby several different integrals of the form
as given in eq.\ (\ref{15}) have been evaluated.  The here presented example
served as decisive test for the achieved accuracy. Given the general features of
the problem we are confident that the algorithm presented here will be valuable
also for other investigations.

\section*{Acknowledgments}
\noindent
We thank {M.Q. Huber} for helpful discussions.\\
Furthermore, we acknowledge support by the 
\textit{Research Core Area ``Modeling and Simulation''} of the 
University of Graz. 
\newpage
\bibliographystyle{model1-num-names}

\end{document}